\documentclass[twocolumn,aps,prl,superscriptaddress]{revtex4-2}

\usepackage{graphicx}
\usepackage[version=3]{mhchem}
\usepackage{braket}
\usepackage{ulem}
\usepackage{here}
\usepackage{dcolumn}
\usepackage{color}
\usepackage{wrapfig}
\usepackage{bm}
\usepackage{xcolor}
\usepackage{lipsum}
\usepackage{cancel}
\usepackage{soul}
\usepackage[utf8]{inputenc}
\usepackage{amsmath}

\begin{document}
\title{Type-II-like ultrafast demagnetization behavior in NiCo$_2$O$_4$ thin films}
\author{Ryunosuke Takahashi}  \email{bagdiners@gmail.com}
\affiliation{Department of Material Science, Graduate School of Science, University of Hyogo, Ako, Hyogo 678-1297, Japan}
\author{Kaede Yamada}
\affiliation{Department of Material Science, Graduate School of Science, University of Hyogo, Ako, Hyogo 678-1297, Japan}
\author{Harjinder Singh}
\affiliation{Institut Jean Lamour, CNRS UMR 7198, Université de Lorraine, F-54506 Nancy, France.}
\author{Kanata Watanabe}
\affiliation{Department of Electrical, Electronics and Information Engineering, Nagaoka University of Technology, Nagaoka 940-2188, Japan}
\author{Junta Igarashi}
\affiliation{Institut Jean Lamour, CNRS UMR 7198, Université de Lorraine, F-54506 Nancy, France.}
\author{Julius Hohlfeld}
\affiliation{Institut Jean Lamour, CNRS UMR 7198, Université de Lorraine, F-54506 Nancy, France.}
\author{Jon Gorchon}
\affiliation{Institut Jean Lamour, UMR CNRS 7198, Universit\'{e} de Lorraine,
BP 70239, F-54506, Vandoeuvre-l\'{e}s-Nancy, France}
\author{Gr\'{e}gory Malinowski}
\affiliation{Institut Jean Lamour, CNRS UMR 7198, Université de Lorraine, F-54506 Nancy, France.}
\author{Daisuke Kan}
\affiliation{Institute for Chemical Research, Kyoto University, Uji, Kyoto 611-0011, Japan}
\author{Yuichi Shimakawa}
\affiliation{Institute for Chemical Research, Kyoto University, Uji, Kyoto 611-0011, Japan}
\author{Takayuki Ishibashi}
\affiliation{Department of Electrical, Electronics and Information Engineering, Nagaoka University of Technology, Nagaoka 940-2188, Japan}
\author{St\'{e}phane Mangin}
\affiliation{Institut Jean Lamour, CNRS UMR 7198, Université de Lorraine, F-54506 Nancy, France.}
\affiliation{Center for Science and Innovation in Spintronics, Tohoku University, Sendai, Japan}
\author{Hiroki Wadati}
\affiliation{Department of Material Science, Graduate School of Science, University of Hyogo, Ako, Hyogo 678-1297, Japan}
\affiliation{Institute of Laser Engineering, Osaka University, Suita, Osaka 565-0871, Japan}

\begin{abstract}
Rare-earth-free ferrimagnetic oxides are emerging as attractive platforms for investigating ultrafast spin dynamics. Here, we study the photoinduced magnetization dynamics of epitaxial NiCo$_2$O$_4$ (NCO) thin films by time-resolved magneto-optical Faraday effect using two independent pump–probe configurations: 1030/515~nm and 800/400~nm. In both measurements, photoexcitation induces an immediate reduction of the magneto-optical signal within the experimental time resolution, followed by a reproducible slower demagnetization component with a characteristic timescale of approximately 5--6~ps and a subsequent recovery on the $\sim 100$~ps timescale. Importantly, this picosecond demagnetization component is observed consistently across the two experimental configurations and excitation wavelengths, demonstrating that it is an intrinsic feature of the ultrafast magnetic response of NCO thin films. Because the earliest-time dip may contain a transient optical contribution, we describe the overall response as type-II-like, rather than assigning a definitive textbook type-II classification solely on the basis of the sub-resolution signal. These results establish a robust two-step ultrafast demagnetization behavior in NCO and highlight rare-earth-free oxide ferrimagnets as promising systems for exploring multisublattice spin dynamics on ultrafast timescales.

\end{abstract}

\maketitle
Since the discovery of ultrafast demagnetization in Ni in 1996 \cite{Beaurepaire1996-ph}, where demagnetization times of less than 1~ps were reported, the underlying microscopic mechanisms have been extensively studied because of their high potential for next-generation non-volatile spintronic applications \cite{Kirilyuk2010-ue,El_Hadri2017-zy}.
Ultrafast demagnetization has been investigated in a wide variety of materials over the past decades \cite{Hohlfeld2000ElectronLattice,Mueller2009HalfMetal,Zhang2006CrO2,Kise2000Sr2FeMoO6,Ogasawara2005PhotoSpin,Gong2023FeGe,Panda2023Permalloy,Legare2024CoMultilayer,Wu2024ThreeStageFGT}, and it is now well established that femtosecond-laser-induced demagnetization exhibits distinctly different temporal characteristics depending on the material system.

Ultrafast demagnetization dynamics have been classified into type-I and type-II behavior depending on the efficiency of angular momentum transfer during electron--phonon equilibration \cite{Koopmans2010}. Type-I dynamics typically exhibit a single-step demagnetization, whereas type-II dynamics are characterized by a two-step process consisting of an initial ultrafast reduction followed by a slower secondary demagnetization. This classification successfully accounts for the contrasting timescales observed across different magnetic materials, such as the sub-picosecond response in transition metals and the much slower dynamics in rare-earth-based systems, including Gd \cite{Koopmans2010}. Moreover, type-II behavior has been reported in several magnetic oxides and multi-sublattice systems, highlighting the importance of material-specific spin dynamics beyond simple ferromagnets.

Most studies of ultrafast demagnetization have been performed on metallic materials, motivating the exploration of ultrafast spin dynamics in rare-earth-free materials composed of light magnetic elements as a promising route toward lightweight and sustainable spintronic applications. In recent years, NiCo$_2$O$_4$ (NCO) thin films have attracted considerable attention as rare-earth-free oxide materials exhibiting perpendicular magnetic anisotropy (PMA) \cite{Xu2022-oi,Kan2020-qo,Kan2020prb,Dho2022-bd}. PMA is of particular interest owing to its potential for high-density magnetic recording.

NCO thin films are ferrimagnetic oxides with a Curie temperature ($T_{\mathrm{C}}$) higher than 400~K for thicknesses of about 30~nm \cite{Shen2020,Shen2020-ml}. In NCO, Co ions occupy both tetrahedral ($T_{d}$) and octahedral ($O_{h}$) sites, while Ni ions occupy the $O_{h}$ sites. X-ray absorption spectroscopy and x-ray magnetic circular dichroism measurements have revealed that the valence states are Co$^{2+}$ at the $O_{h}$ sites, Co$^{3+}$ at the $T_{d}$ sites, and Ni$^{2+\delta}$ at the $O_{h}$ sites \cite{Kan2020prb}. The spins of the $T_{d}$-site Co and $O_{h}$-site Ni are ferrimagnetically coupled \cite{Bitla2015,Kan2020prb}, resulting in a saturated magnetization of approximately 2~$\mu_{\mathrm{B}}$/f.u. \cite{Kan2020-qo}.
Recent studies by our group have demonstrated that NCO thin films exhibit all-optical switching at room temperature and at elevated temperatures close to $T_c$, accompanied by rich light-induced magnetic phenomena resulting from a complex interplay of magnetic circular dichroism effects and stray fields \cite{Takahashi2023,Takahashi2025NCO_AOS}. Our previous studies have also reported ultrafast demagnetization via time-resolved magneto-optical Kerr effect microscopy in NCO thin films \cite{Takahashi2021,Shen2020-ml}. 

While these results establish NCO as a promising material, the detailed temporal evolution of the demagnetization process and its classification within the type-I/type-II framework remain largely unexplored. The metallic electronic state of NCO thin films, confirmed by valence-band XPS measurements \cite{Takahashi2021}, suggests that electron–lattice–mediated spin-flip scattering could drive a type-I-like ultrafast demagnetization. On the other hand, as a magnetic oxide hosting multiple magnetic sublattices, NCO may also exhibit features characteristic of type-II-like dynamics. Consequently, the ultrafast magnetic response of this compound is nontrivial. Establishing the dynamical type in NCO would clarify whether its ultrafast response is governed primarily by electron–phonon–mediated spin-flip scattering or by more complex sublattice-dependent magnetic interactions.

In this study, we investigate the ultrafast demagnetization dynamics in thin films of the rare-earth-free ferrimagnetic oxide NCO using time-resolved magneto-optical Faraday effect (TR-MOFE) measurements. Magnetization dynamics induced by femtosecond laser excitation at 1030 and 800 nm were examined using two independent experimental setups to clarify the temporal evolution of the demagnetization process and to determine its classification within the type-I/type-II framework. Particular attention is paid to the two-stage demagnetization process, consisting of an ultrafast magnetization reduction occurring within the experimental time resolution, followed by a picosecond-scale slower component and a subsequent recovery process, in order to elucidate the origin of the type-II-like ultrafast demagnetization behavior in NCO thin films.

Epitaxial NCO thin films with a thickness of $\sim$30~nm were grown on MgAl$_2$O$_4$ (001) substrates by pulsed laser deposition. During film growth, the substrate temperature and oxygen partial pressure were maintained at 315$^\circ$C and 100~mTorr, respectively. Details of the NCO thin-film growth procedure are described in Refs.~\cite{Kan2020-qo,Shen2020}.
\begin{figure}[t]
  \centering
  \includegraphics[width=\linewidth]{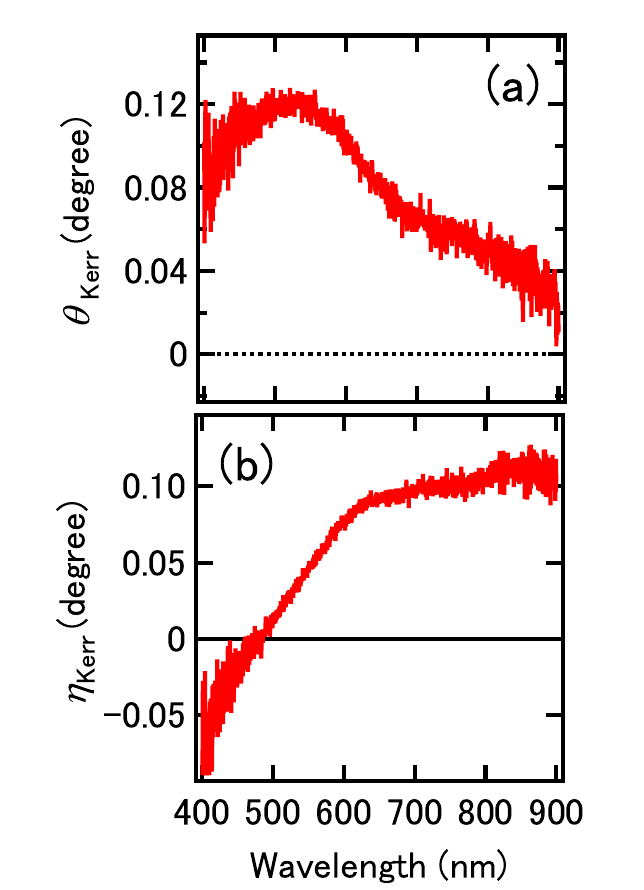}
  \caption{Magneto-optical Kerr effect MOKE spectrum of the NCO thin film.
  Panel (a) shows the real part of the Kerr response, corresponding to the Kerr rotation angle $\theta$, while panel (b) displays the imaginary part, corresponding to the Kerr ellipticity $\varepsilon$.
  The Kerr rotation reaches a maximum of approximately 0.13~deg at around 521~nm.
  The Kerr ellipticity changes sign near 482~nm, indicating a dispersive magneto-optical response within the visible spectral range.}
  \label{fig:moke_spectrum}
\end{figure}

Magneto-optical Kerr effect (MOKE) spectra were measured to investigate the magneto-optical effect of NCO thin films. The polar magneto-optical Kerr effect (p-MOKE) spectra were obtained in a polar configuration using a MOKE spectrometer~\cite{Sato1993,Sakaguchi2023Mn4NB}. The measurements were performed with a broadband optical setup. Further details of the experimental setup are described in Ref.~\cite{Sakaguchi2023Mn4NB}.

\begin{figure*}[t]
  \centering
  \includegraphics[width=\textwidth]{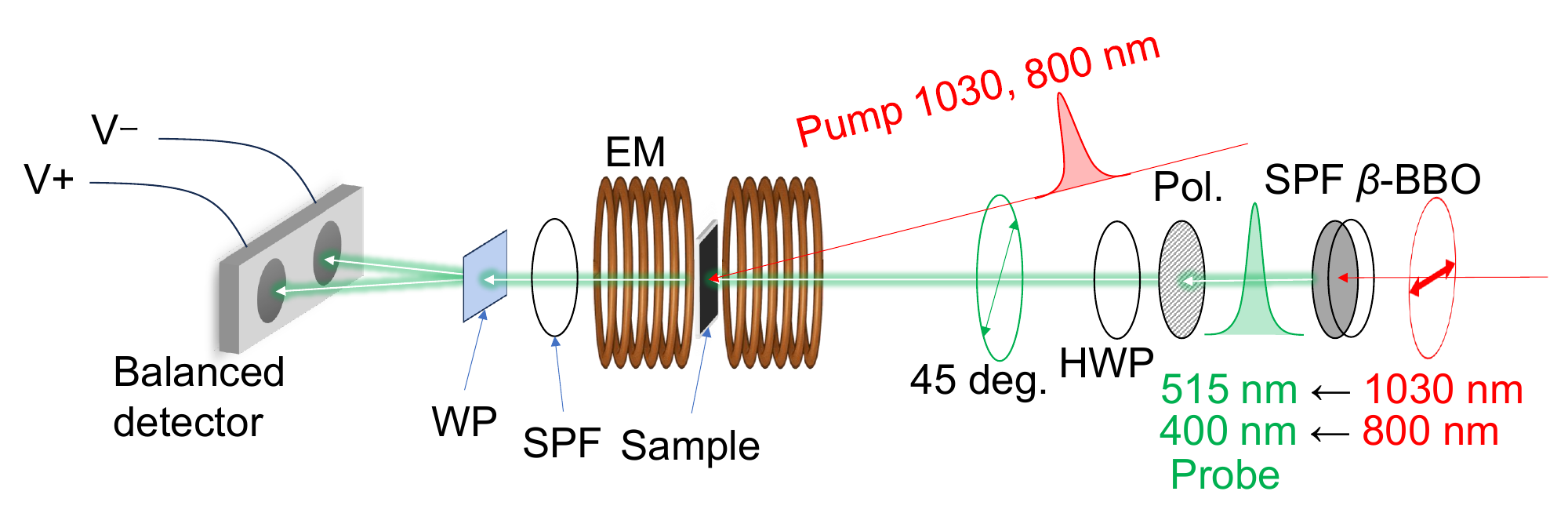}
  \caption{Schematic illustration of the time-resolved magneto-optical Faraday effect measurement setup used in the 1030/515~nm and 800/400~nm pump--probe experiments.
  A femtosecond pump pulse at 1030~nm or 800 nm excites the sample, while the probe pulse at 515~nm or 400 nm,
  generated by second-harmonic generation, monitors the transient Faraday rotation.
  The time delay between the pump and probe pulses is controlled by a mechanical delay line.
  The polarization rotation of the transmitted probe beam is detected using a balanced
  photodetector. WP: Wollaston prism; HWP: half-wave plate; DM: dichroic mirror; SPF: short-pass
  filter; EM: electromagnet.}
  \label{fig:setup}
\end{figure*}

Figure~\ref{fig:moke_spectrum} shows the MOKE spectrum of the NCO thin film measured in the wavelength range of 400–900~nm.
The magneto-optical Kerr rotation $\theta(\lambda)$ exhibits a broad maximum at approximately 521~nm, where the rotation angle reaches about 0.13~deg.
Rather than forming a sharp spectral line, the feature extends over several tens of nanometers.
In addition, the Kerr ellipticity $\varepsilon(\lambda)$ changes sign at approximately 482~nm.
The coexistence of a broad rotational maximum and a zero crossing in the ellipticity reflects the dispersive character of the complex magneto-optical Kerr response across an electronic resonance.
Such a line shape is consistent with magneto-optical enhancement associated with a wide interband or charge-transfer transition, rather than a single narrow crystal-field ($d$–$d$) excitation.

Additionally, prior to the time-resolved measurements, the magnetic-field dependence of the magneto-optical Faraday rotation angle ($\theta_{\mathrm{F}}$) was measured using a 515 nm probe beam generated as the second harmonic of PHAROS to calibrate $\theta_{\mathrm{F}}$.
$\theta_{\mathrm{F}}$ was evaluated using a balanced detection scheme with a Wollaston prism, as illustrated in Fig.~\ref{fig:setup}.

\begin{figure}[t]
  \centering
  \includegraphics[width=\linewidth]{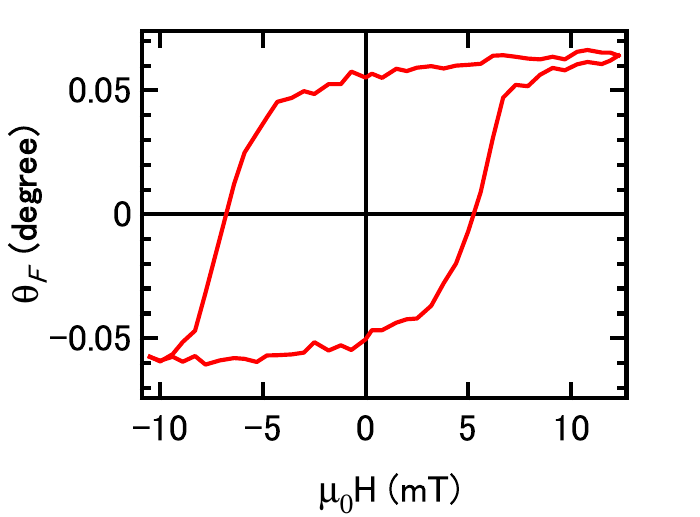}
  \caption{Magnetic-field dependence of the magneto-optical Faraday rotation angle $\theta_{\mathrm{F}}$ of the NiCo$_2$O$_4$ thin film at 515 nm probe.
  $\theta_{\mathrm{F}}$ saturates at approximately $\pm 60$~mdeg at magnetic fields on the order of 10~mT. The coercive field is approximately 6~mT.}
  \label{fig:faraday_field}
\end{figure}

Figure~\ref{fig:faraday_field} shows the magnetic-field dependence of $\theta_{\mathrm{F}}$ measured prior to the time-resolved experiments. The Faraday rotation saturates at approximately $\pm 60$~mdeg at magnetic fields around 10~mT. The coercive field, estimated from the hysteresis loop, is approximately 6~mT. This value is obtained as the average of the magnetic fields at which the magnetization crosses zero during positive and negative field sweeps.

TR-MOFE measurements were performed using a pump--probe technique. At the University of Hyogo, a linearly polarized femtosecond laser pulse with a wavelength of 1030~nm was used as the pump to excite the sample, and the transient magneto-optical response was probed using a 515~nm pulse generated via second-harmonic generation.

Measurements were also carried out at the Institut Jean Lamour (IJL) in Nancy, France, using a Ti:sapphire laser system. In this setup, an 800~nm pump pulse and a 400~nm probe pulse generated via second-harmonic generation were employed, and the $\theta_{\mathrm{F}}$ signal was detected using a lock-in amplifier. The pulse duration was approximately 50~fs, and the time resolution was estimated to be on the order of 70~fs, as determined from the cross-correlation between the pump and probe pulses.

$\theta_{\mathrm{F}}$ signal was measured as a function of the pump--probe time delay.
To extract the magnetization-related contribution, the TR-MOFE signals were measured under magnetic fields of $+\mu_0 H$ and $-\mu_0 H$. 
The asymmetric component with respect to magnetic field reversal was calculated as
\begin{equation}
\theta_{\mathrm{asym}}(t)
=
\frac{\theta(+\mu_0 H,t)-\theta(-\mu_0 H,t)}{2},
\end{equation}
where $\theta(+\mu_0 H,t)$ and $\theta(-\mu_0 H,t)$ denote the TR-MOFE signals measured under positive and negative magnetic fields, respectively. 
All measurements were carried out at room temperature in a Faraday geometry, with the magnetic field applied perpendicular to the film plane. 
The external magnetic field was generated by an electromagnet. 
In both the University of Hyogo and the IJL experiments, the measurements were performed under magnetic fields exceeding the saturation field of the sample. 
More detailed experimental procedures are described in Appendix~A.

Figure~\ref{fig:trfr} shows the TR-MOFE signals of the NCO thin film. Immediately after photoexcitation, a rapid decrease in $\theta_{\mathrm{F}}$ is observed within the experimental time resolution, indicating an ultrafast demagnetization process. This initial response is followed by a slower demagnetization component on a picosecond timescale and a subsequent recovery process extending to hundreds of picoseconds.

Panels (a) and (b) display the results obtained using the setup at the University of Hyogo. Panel (a) shows the TR-MOFE signal over a wide time window from $-2$ to 106~ps, capturing both the initial demagnetization and the subsequent recovery. Panel (b) presents an enlarged view of the early-time dynamics from $-2$ to 20~ps, highlighting the fast demagnetization immediately after photoexcitation.

Panels (c) and (d) show the corresponding measurements performed at the IJL. Panel (c) extends the observation window up to approximately 500~ps, providing access to the long-time recovery dynamics, while panel (d) focuses on the early-time response within the first 40~ps. Despite differences in the pump and probe wavelengths, the overall temporal evolution of the magnetization is consistent across the two independent measurement setups.

The experimental data were analyzed by fitting the TR-MOFE traces with a multi-component model consisting of fast and slow demagnetization components and a recovery component. The explicit form of the fitting function and the fitting procedure are described in Appendix B.

The initial dip may be influenced by a transient modification of the magneto-optical response associated with photoexcited carriers and therefore does not necessarily reflect a pure change in magnetization~\cite{Guidoni2002}. In this study, the magnetization-related signal was extracted as the antisymmetric component with respect to magnetic field reversal. Nevertheless, a small contribution from transient optical effects may remain, for example due to imperfect cancellation of field-symmetric components. 
This feature is reproducible across different experimental setups, increases with pump fluence, and exhibits a temporal width of approximately 0.3~ps, which is significantly larger than the instrumental response. Similar behavior has also been reported in other oxide systems, such as La$_{1-x}$Sr$_x$MnO$_3$~\cite{Mueller2009HalfMetal}. These observations indicate that the dip reflects an intrinsic ultrafast process rather than a convolution-limited response. While such behavior is consistent with ultrafast magnetization dynamics, a contribution from magnetic-field-dependent transient optical effects associated with photoexcited carriers—such as transient magnetic circular dichroism or magnetic birefringence—cannot be excluded. 
For this reason, in order to evaluate the intrinsic magnetic dynamics of the present material, the dip structure immediately after $t=0$~ps was excluded from the fitting range (0.1--0.4~ps).

For the measurements performed at the University of Hyogo, the fitting yielded three characteristic time constants. 
The experimental data indicate that the initial demagnetization process occurs on a timescale shorter than the instrumental time resolution. When treated as a free parameter, $\tau_{\mathrm{fast}}$ collapsed to the lower bound of the fitting range ($\sim 0.001$~ps), confirming that it cannot be resolved within the present temporal resolution.
In addition, a slower demagnetization process with a time constant of $\tau_{\mathrm{slow}} = 5.59$~ps was clearly resolved. 
The recovery of the magneto-optical Faraday rotation signal exhibits a long relaxation time, and the recovery time constant $\tau_{\mathrm{rec}}$ could not be reliably determined within the experimental time window. Although the fitting yields a value of $56.1 \pm 15.47$ ps, the large uncertainty reflects the limitation of the temporal window. Therefore, the recovery process cannot be quantitatively discussed based on the present data. 

In the IJL setup, measurements were performed over a wider temporal window and for a broader range of excitation fluences, and the same fitting procedure was applied. 
The detailed fitting results are summarized in the Appendix~B.

In both experimental setups, qualitatively identical dynamics were observed, consisting of a two-step demagnetization followed by a single-exponential recovery. 
The first demagnetization step occurs on a timescale shorter than the experimental time resolution (70~fs), preventing a reliable determination of its intrinsic time constant.
The second demagnetization process exhibits a characteristic time constant of approximately $\tau_{\mathrm{slow}} \sim 5$--6~ps over the investigated fluence range, showing no systematic dependence on excitation fluence. 

In contrast, the recovery time exhibits a clear fluence dependence. At an excitation fluence of 0.10~mJ/cm$^2$, the recovery time is approximately 95~ps, increasing to about 224~ps at 0.25~mJ/cm$^2$, and remaining similarly long at 0.40~mJ/cm$^2$. 
These results indicate that the recovery dynamics become significantly slower above 0.10~mJ/cm$^2$, while no further systematic change is resolved within the experimental uncertainty at higher fluences. 
This behavior may reflect enhanced lattice heating, which reduces the effective driving force for spin–lattice relaxation and leads to a bottleneck in angular momentum transfer at higher excitation fluences.
\begin{figure}[t]
  \centering
  \includegraphics[width=\linewidth]{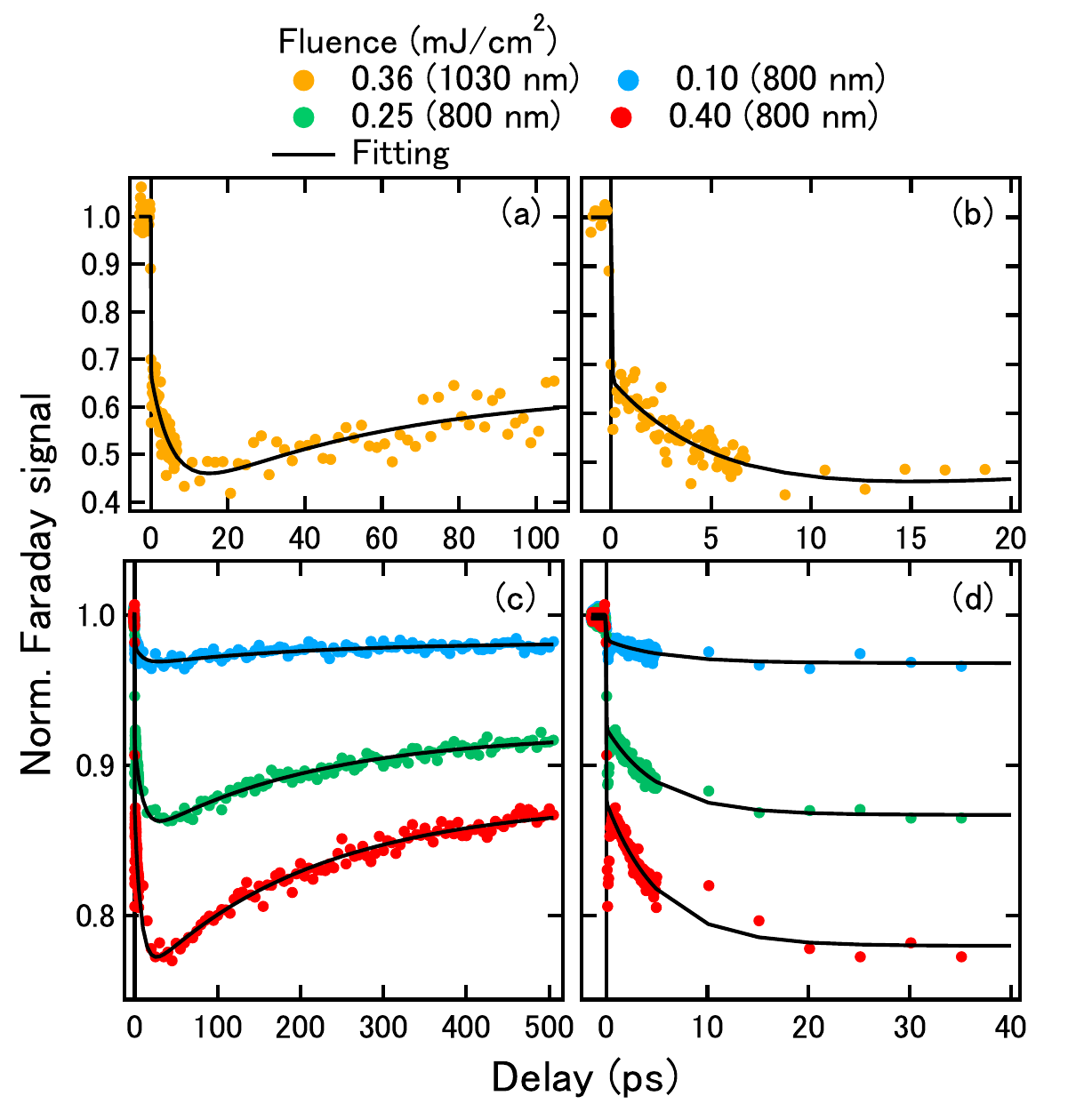}
  \caption{Time-resolved magneto-optical Faraday effect (TR-MOFE) of the NiCo$_2$O$_4$ thin film following femtosecond laser excitation at 1030 and 800~nm. The symbols represent the experimental data, and the solid line is a fit using a multi-component model consisting of an ultrafast demagnetization component, a slower picosecond-scale demagnetization component, and a recovery component. Panels (a) and (b) show the results obtained using the present setup: (a) displays the data over a long time window from $-2$ to 106~ps, while (b) shows an expanded view in the short time range from $-2$ to 20~ps. Panels (c) and (d) present additional measurements performed using the TR-MOFE apparatus at Institut Jean Lamour, demonstrating the pump-fluence dependence measured at 0.10, 0.25, and 0.40~mJ/cm$^{2}$; panel (c) shows the data up to 500~ps, and panel (d) displays an enlarged view of the early-time dynamics within the first 40~ps. The values on the right of the fluence denote the pump wavelength.}
  \label{fig:trfr}
\end{figure}
Additionaly, our previous studies reported a demagnetization timescale of about 0.4~ps \cite{Takahashi2021}; however, the characteristic timescale obtained in the present study is different. This discrepancy may be attributed to the data analysis procedure, especially the use of a single-exponential fitting model, which may mask multiple relaxation components in the demagnetization dynamics.

\section{Discussion}

Koopmans \textit{et al.} proposed the ratio $T_C/\mu_{at}$ as a figure of merit for ultrafast demagnetization dynamics, predicting that materials with smaller $T_C/\mu_{at}$ tend to exhibit slower demagnetization and a characteristic two-step (type-II) response, assuming comparable spin-flip scattering probabilities \cite{Koopmans2010}.

Applying this classification to NCO thin films, we treat $\mu_{at}$ as the average magnetic moment per magnetic transition-metal ion, i.e., per Ni and Co atom, consistent with the definition of the atomic magnetic moment used by Koopmans \textit{et al.} Using the low-temperature saturated magnetic moment $\mu_{\mathrm{f.u.}} \approx 2~\mu_B/\mathrm{f.u.}$ and $T_C \approx 400$~K, we obtain $\mu_{at} = \mu_{\mathrm{f.u.}}/3 \approx 0.67~\mu_B$ and $T_C/\mu_{at} \approx 600~\mathrm{K}/\mu_B$. 
For comparison, typical type-I transition metals such as Ni and Co exhibit $T_C/\mu_{at} \sim 800$--$1000~\mathrm{K}/\mu_B$, whereas rare-earth systems such as Gd show much smaller values of $T_C/\mu_{at} \sim 40~\mathrm{K}/\mu_B$, corresponding to type-II dynamics. 
The value for NCO therefore lies in an intermediate regime between these two limits. However, since NCO is an oxide with magnetic sublattices, the possibility of type-II behavior cannot be excluded.

Despite this boundary-like value of $T_C/\mu_{at}$, the present TR-MOFE measurements clearly reveal a two-step demagnetization behavior, which is a hallmark of type-II-like ultrafast magnetization dynamics. This behavior can be naturally understood in terms of the oxide nature of NCO, characterized by multiple magnetic sublattices. In addition, the spin polarization at room temperature is finite (approximately 0.7), indicating an incomplete half-metallic character and the presence of minority-spin states near the Fermi level. Although NCO exhibits metallic charge transport, its magnetic moments originate from partially localized $3d$ electrons, and their coupling to optically excited carriers is therefore weaker and less direct than in elemental itinerant ferromagnets.

Furthermore, NCO has an inverse spinel structure with mixed valence states of Ni and Co and multiple magnetic sublattices, which can naturally give rise to distinct magnetic relaxation channels with different characteristic timescales. These oxide-specific features provide a plausible microscopic basis for the emergence of type-II-like dynamics, even though the static figure of merit $T_C/\mu_{at}$ places the material near the type-I/type-II boundary. Indeed, type-II ultrafast demagnetization behavior has been reported for several complex oxides, including CrO$_2$\cite{Zhang2006CrO2,Mueller2009HalfMetal}, Sr$_2$FeMoO$_6$\cite{Kise2000Sr2FeMoO6}, and La$_{1-x}$Sr$_x$MnO$_3$\cite{Mueller2009HalfMetal,Ogasawara2005PhotoSpin}.

Another important point is that, although Co$_2$MnAl$_x$Si$_{1-x}$ and La$_{1-x}$Sr$_x$MnO$_3$ are both nearly fully spin-polarized systems, their demagnetization timescales differ by orders of magnitude, ranging from a few hundred femtoseconds to several hundred picoseconds\cite{Guillemard2020,Mueller2009HalfMetal}. This clearly indicates that the spin polarization alone is not the dominant factor determining the demagnetization timescale. To understand ultrafast magnetization dynamics in oxide systems such as La$_{1-x}$Sr$_x$MnO$_3$ and NCO, it is essential to consider additional factors, including electron correlation, spin–lattice coupling, and multi-sublattice magnetism.

Taken together, the boundary value of $T_C/\mu_{at}$, the oxide-specific electronic and magnetic structure, and the direct observation of two-step demagnetization at room temperature consistently indicate that NCO thin films exhibit type-II-like ultrafast magnetization dynamics. 
This conclusion further suggests that the ultrafast demagnetization in NCO cannot be explained solely by electron–phonon–mediated spin-flip scattering, but is instead largely governed by multi-sublattice magnetic interactions characteristic of oxide ferrimagnets.

In summary, we have investigated the ultrafast magnetization dynamics of epitaxial NCO thin films by TR-MOFE using two independent experimental configurations with different pump and probe wavelengths. In both cases, we observe the same qualitative sequence: an immediate drop in the magneto-optical signal within the experimental time resolution, a slower demagnetization component with a characteristic timescale of approximately 5--6~ps, and a much slower recovery extending over tens to hundreds of picoseconds. The reproducible observation of the 5--6~ps component in both the 1030/515~nm and 800/400~nm measurements is a central result of this work, because it demonstrates that the slower demagnetization is not a setup-specific artifact but an intrinsic property of NCO.

At the same time, the earliest sub-resolution dip appears in the same temporal window as a pump-induced optical response and may therefore contain a nonmagnetic contribution. For this reason, the present data support a cautious classification in terms of type-II-like ultrafast demagnetization. Taken together, our results show that NCO exhibits a robust two-step ultrafast magnetic response, despite its location near the type-I/type-II boundary inferred from the $T_\mathrm{C}/\mu_{\mathrm{at}}$ criterion. More broadly, this work extends the discussion of type-II-like demagnetization to rare-earth-free oxide ferrimagnets and underlines the importance of oxide-specific electronic structure and multisublattice magnetism in shaping ultrafast spin dynamics.

\section{acknowledgments}
The MOKE spectrometer used in this study was provided by the Department of Materials Science and Technology, Nagaoka University of Technology. This work was supported by the Institut Carnot ICEEL, the Région Grand Est, and the Métropole du Grand Nancy through the project ``OPTIMAG.'' It was also supported by the Agence Nationale de la Recherche (ANR) through the SLAM project (ANR-23-CE30-0047), as well as by the France 2030 government programs, including PEPR Electronic EM-COM (ANR-22-PEEL-0009), PEPR SPIN (SPINMAT, ANR-22-EXSP-0007), and Chirex (ANR-22-EXSP-0002). Additional support was provided by the MAT-PULSE Lorraine University d’Excellence project (ANR-15-IDEX-04-LUE).

This article is based on work from COST Action CA23136 (CHIROMAG), supported by COST (European Cooperation in Science and Technology). This work was also supported by JSPS KAKENHI under Grant Nos.~23H01108, 23K25805, and JP25H01251, and by the MEXT Quantum Leap Flagship Program (MEXT Q-LEAP) under Grant No.~JPMXS0118068681. Further support was provided by the Adopting Sustainable Partnerships for Innovative Research Ecosystem (ASPIRE), Grant No.~JPMJAP2314, from the Japan Science and Technology Agency (JST). This work was partly supported by grants for the Integrated Research Consortium on Chemical Sciences and the International Collaborative Research Program of the Institute for Chemical Research, Kyoto University, from Japan's Ministry of Education, Culture, Sports, Science, and Technology (MEXT).

The data that support the findings of this study are available from the corresponding author upon reasonable request.

\appendix

\setcounter{equation}{0}
\setcounter{figure}{0}
\setcounter{table}{0}

\renewcommand{\theequation}{A.\arabic{equation}}
\renewcommand{\thefigure}{A.\arabic{figure}}
\renewcommand{\thetable}{A.\arabic{table}}

\section*{Appendix A: Details of TR-MOFE Measurements}

The time delay between the pump and probe pulses was controlled by a mechanical delay line in the pump beam path. 

The probe polarization was adjusted using a half-wave plate to achieve a balanced condition ($V_1 = V_2$) in the absence of magnetization, corresponding to a linearly polarized state at $45^\circ$. Residual pump light was removed prior to detection by a short-pass filter, as shown in Fig.~\ref{fig:setup}.

The probe beam transmitted through the sample was separated into two orthogonally polarized components using a Wollaston prism, and the magneto-optical Faraday rotation was detected by a balanced photodetector. The intensities of the two components were simultaneously measured as $V_1$ and $V_2$, which are given by
\begin{equation}
V_1 = V_0 \cos^2 \left(\pi/4 + \theta_{\mathrm{F}}\right), \qquad
V_2 = V_0 \sin^2 \left(\pi/4 + \theta_{\mathrm{F}}\right),
\label{eq:faraday_signal}
\end{equation}
where $V_0$ is proportional to the probe intensity, and $\theta_{\mathrm{F}}$ denotes the magneto-optical Faraday rotation angle.

The normalized differential signal was defined as
\begin{equation}
S = \frac{V_2 - V_1}{V_1 + V_2},
\label{eq:balance_signal}
\end{equation}

For small $\theta_{\mathrm{F}} \ll 1$, this relation can be approximated as $S = \sin(2\theta_{\mathrm{F}}) \simeq 2\theta_{\mathrm{F}}$, allowing $\theta_{\mathrm{F}}$ to be directly obtained from the measured differential signal.

\setcounter{equation}{0}
\setcounter{figure}{0}
\setcounter{table}{0}

\renewcommand{\theequation}{B.\arabic{equation}}
\renewcommand{\thefigure}{B.\arabic{figure}}
\renewcommand{\thetable}{B.\arabic{table}}

\section*{Appendix B: Fitting Procedure and Extracted Parameters}

\begin{table*}[t]
\centering
\caption{Fluence dependence of the fitting parameters.}
\begin{tabular}{c|cc|cc|cc}
\hline
 & \multicolumn{2}{c|}{Fast} 
 & \multicolumn{2}{c|}{Slow} 
 & \multicolumn{2}{c}{Recovery} \\
\cline{2-7}
Fluence 
& $A_{\mathrm{fast}}$ & $\tau_{\mathrm{fast}}$ (ps)
& $A_{\mathrm{slow}}$ & $\tau_{\mathrm{slow}}$ (ps)
& $A_{\mathrm{rec}}$  & $\tau_{\mathrm{rec}}$ (ps) \\
(mJ/cm$^{2}$) & & & & & & \\
\hline
0.10 
& $-0.017 \pm 0.001$ & $< 0.07$
& $-0.015 \pm 0.002$ & $5.7 \pm 1.20$
& $0.012 \pm 0.001$  & $95.2 \pm 22$ \\

0.25 
& $-0.073 \pm 0.0014$ & $< 0.07$
& $-0.058 \pm 0.002$ & $5.1 \pm 0.37$
& $0.059 \pm 0.0020$  & $223.6 \pm 22$ \\

0.36 
& $-0.33 \pm 0.013$  & $< 0.3$
& $-0.29 \pm 0.036$  & $5.59 \pm 0.35$
& $0.26 \pm 0.036$   & $56.1 \pm 15.47$ \\

0.40 
& $-0.123 \pm 0.0019$ & $< 0.07$
& $-0.096 \pm 0.0028$ & $5.35 \pm 0.32$
& $0.107 \pm 0.0026$  & $223.6 \pm 16$ \\
\hline
\end{tabular}
\end{table*}

The transient signal was analyzed using a multi-component model function,
\begin{align}
y_{\mathrm{fit}}(t)
&= C + \Bigl[ H(t-t_0)\,
\bigl( s_{\mathrm{fast}}(t-t_0) \nonumber \\
&\quad + s_{\mathrm{slow}}(t-t_0)
+ s_{\mathrm{rec}}(t-t_0) \bigr) \Bigr]
* \mathrm{IRF}(t),
\end{align}
where $H(t)$ is the Heaviside step function. The instrumental response function $\mathrm{IRF}(t)$ was modeled by a Gaussian function,
\begin{equation}
\mathrm{IRF}(t)
= \frac{1}{\sqrt{2\pi}\sigma}
\exp\left(-\frac{t^{2}}{2\sigma^{2}}\right),
\end{equation}
where $\sigma$ is related to the full width at half maximum (FWHM) of the temporal resolution. The value of $\sigma$ was fixed according to the time resolution of each experimental setup: $\sigma = 0.3~\mathrm{ps}$ for the University of Hyogo measurements and $\sigma = 70~\mathrm{fs}$ for the IJL measurements. The time zero $t_0$ was treated as a free fitting parameter.

Each dynamical component was described by
\begin{equation}
s_i(t) = A_i \left(1 - e^{-t/\tau_i}\right),
\qquad
i \in \{\mathrm{fast}, \mathrm{slow}, \mathrm{rec}\},
\end{equation}
where $A_i$ and $\tau_i$ denote the amplitude and time constant of each component, respectively.

To estimate the demagnetization time constant, the dip structure immediately after $t=0$~ps was excluded from the fitting range. Specifically, the data in the range of 0.1--0.4~ps were masked.
Table~B.1 summarizes the fitting parameters obtained from the TR-MOFE measurements.
The fast time constant $\tau_{\mathrm{fast}}$ was treated as a free parameter but collapsed to the lower bound of the fitting range ($\sim 0.001$~ps), indicating that it is well below the instrumental response and thus cannot be resolved. Accordingly, in Table~B.1, the fast component is noted as being below the temporal resolution. The amplitudes and the other time constants were treated as free parameters. The fitting was performed using a nonlinear least-squares method, and the quoted uncertainties correspond to one standard deviation ($1\sigma$). 

\bibliography{references}

@article{Sato1993,
  author  = {Sato, Katsuaki and Hongu, Hidetoshi and Ikekame, Hiroshi and Tosaka, Yasuhiro and Watanabe, Masato and Takanashi, Koki and Fujimori, Hiroyasu},
  title   = {Magnetooptical Kerr Spectrometer for 1.2--5.9 eV Region and its Application to {FePt/Pt} Multilayers},
  journal = {Jpn.\ J.\ Appl.\ Phys.},
  volume  = {32},
  number  = {2R},
  pages   = {989--994},
  year    = {1993},
  doi     = {10.1143/JJAP.32.989},
  publisher = {The Jpn. Soc. Appl. Phys.}
}

@article{Koopmans2010,
  author  = {Koopmans, B. and Malinowski, G. and Dalla Longa, F. and Steiauf, D. and F{\"a}hnle, M. and Roth, T. and Cinchetti, M. and Aeschlimann, M.},
  title   = {Explaining the paradoxical diversity of ultrafast laser-induced demagnetization},
  journal = {Nat.\ Mater.},
  volume  = {9},
  pages   = {259--265},
  year    = {2010},
  doi     = {10.1038/nmat2593}
}

@article{Sakaguchi2023Mn4NB,
  title   = {Boron-induced magneto-optical Kerr spectra and dielectric tensors in ferrimagnetic {(Mn$_4$N)B} antiperovskite thin films},
  author  = {Sakaguchi, Hotaka and Isogami, Shinji and Niimi, Makoto and Ishibashi, Takayuki},
  journal = {J. Phys. D: Appl. Phys.},
  volume  = {56},
  number  = {36},
  pages   = {365002},
  year    = {2023},
  doi     = {10.1088/1361-6463/acd9d4}
}

@article{Zhang2006CrO2,
  title   = {Ultrafast spin dynamics in half-metallic {CrO$_2$} thin films},
  author  = {Zhang, Q. and Nurmikko, A.~V. and Miao, G.~X. and Xiao, G. and Gupta, A.},
  journal = {Phys. Rev. B},
  volume  = {74},
  pages   = {064414},
  year    = {2006},
  doi     = {10.1103/PhysRevB.74.064414}
}

@article{Kise2000Sr2FeMoO6,
  title   = {Ultrafast spin dynamics and critical behavior in half-metallic ferromagnet {Sr$_2$FeMoO$_6$}},
  author  = {Kise, T. and Ogasawara, T. and Matsubara, M. and Tomioka, Y. and Tokura, Y. and Kuwata-Gonokami, M.},
  journal = {Phys. Rev. Lett.},
  volume  = {85},
  pages   = {1986--1989},
  year    = {2000},
  doi     = {10.1103/PhysRevLett.85.1986}
}

@article{Hohlfeld2000ElectronLattice,
  title   = {Electron and lattice dynamics following optical excitation of metals},
  author  = {Hohlfeld, J. and Wellershoff, S.-S. and G{\"u}dde, J. and Conrad, U. and J{\"a}hnke, V. and Matthias, E.},
  journal = {Chem. Phys.},
  volume  = {251},
  number  = {1--3},
  pages   = {237--258},
  year    = {2000}
}

@article{Guillemard2020,
  author = {Guillemard, Charles and Zhang, Wei and Malinowski, Gregory and de Melo, Claudia and Gorchon, Jon and Petit-Watelot, Sebastien and Ghanbaja, Jaafar and Mangin, Stephane and Le F{\`e}vre, Pierre and Bertran, Franck and Andrieu, Stephane},
  title = {Heusler Alloys for Spintronics: Tailoring Materials for Ultrafast Spin Dynamics},
  journal = {Adv. Mater.},
  volume = {32},
  pages = {1908357},
  year = {2020},
  doi = {10.1002/adma.201908357}
}

@article{Guidoni2002,
  author = {Guidoni, Luca and Beaurepaire, Eric and Bigot, Jean-Yves},
  title = {Magneto-optics in the ultrafast regime: Thermalization of spin populations in ferromagnetic films},
  journal = {Phys. Rev. Lett.},
  volume = {89},
  pages = {017401},
  year = {2002},
  doi = {10.1103/PhysRevLett.89.017401}
}

@article{Legare2024CoMultilayer,
  author = {L{\'e}gar{\'e}, Katherine and Barrette, Guillaume and Giroux, Laurent and Parent, Jean-Michel and Haddad, Elissa and Ibrahim, Heide and Lassonde, Philippe and Jal, Emmanuelle and Vodungbo, Boris and L{\"u}ning, Jan and Boschini, Fabio and Jaouen, Nicolas and L{\'e}gar{\'e}, Fran{\c{c}}ois},
  title = {Near- and mid-infrared excitation of ultrafast demagnetization in a cobalt multilayer system},
  journal = {Phys. Rev. B},
  volume = {109},
  pages = {094407},
  year = {2024},
  doi = {10.1103/PhysRevB.109.094407}
}

@article{Panda2023Permalloy,
  title   = {Ultrafast demagnetization and precession in permalloy films with varying thickness},
  author  = {Panda, S.~N. and Mondal, S. and Majumder, S. and Barman, A.},
  journal = {Phys. Rev. B},
  volume  = {108},
  pages   = {144421},
  year    = {2023},
  doi     = {10.1103/PhysRevB.108.144421}
}

@article{Gong2023FeGe,
  author = {Gong, Zizhao and Zhang, Wei and Liu, Jianan and Xie, Zongkai and Yang, Xu and Tang, Jin and Du, Haifeng and Li, Na and Zhang, Xiangqun and He, Wei and Cheng, Zhao-hua},
  title = {Ultrafast demagnetization dynamics in the epitaxial FeGe(111) film chiral magnet},
  journal = {Phys. Rev. B},
  volume = {107},
  pages = {144429},
  year = {2023},
  doi = {10.1103/PhysRevB.107.144429}
}

@article{Takahashi2025NCO_AOS,
  title   = {All-optical helicity-dependent switching in {NiCo$_2$O$_4$} thin films},
  author  = {Takahashi, R. and Le Guen, Y. and Nakata, S. and Igarashi, J. and Hohlfeld, J. and Malinowski, G. and Xie, L. and Kan, D. and Shimakawa, Y. and Mangin, S. and Wadati, H.},
  journal = {Appl. Phys. Lett.},
  volume  = {126},
  pages   = {212405},
  year    = {2025},
  doi     = {10.1063/5.0253785}
}

@article{Wu2024ThreeStageFGT,
  title   = {Three-stage ultrafast demagnetization dynamics in a monolayer ferromagnet},
  author  = {Wu, N. and Zhang, S. and Chen, D. and Wang, Y. and Meng, S.},
  journal = {Nat. Commun.},
  volume  = {15},
  pages   = {2804},
  year    = {2024},
  doi     = {10.1038/s41467-024-47128-4}
}

@article{Ogasawara2005PhotoSpin,
  title   = {General features of photoinduced spin dynamics in ferromagnetic and ferrimagnetic compounds},
  author  = {Ogasawara, T. and Tomioka, Y. and Matsubara, M. and Ohbayashi, K. and Ueda, S. and Okimoto, Y. and Okamoto, H. and Tokura, Y.},
  journal = {Phys. Rev. Lett.},
  volume  = {94},
  pages   = {087202},
  year    = {2005},
  doi     = {10.1103/PhysRevLett.94.087202}
}

@article{Mueller2009HalfMetal,
  title   = {Spin polarization in half-metal{s} probed by femtosecond spin excitation},
  author  = {M{\"u}ller, G. and Walowski, J. and Djordjevic Kaufmann, S. and Schneider, H. C. and Milovsevic, M. and Bayer, C. and Mali{\'c}, E. and McGuire, M. A. and Kavich, J. J. and Galanakis, I. and Mavropoulos, P. and Bl{\"u}gel, S. and Woltersdorf, G. and Back, C. H. and K{\"o}hler, A. and Martin, T. and Cinchetti, M. and Aeschlimann, M.},
  journal = {Nat. Mater.},
  volume  = {8},
  pages   = {56--61},
  year    = {2009},
  doi     = {10.1038/nmat2357}
}

@ARTICLE{Beaurepaire1996-ph,
  title    = "Ultrafast spin dynamics in ferromagnetic nickel",
  author   = "Beaurepaire, E and Merle, J and Daunois, A and Bigot, J",
  journal  = "Phys. Rev. Lett.",
  volume   =  76,
  number   =  22,
  pages    = "4250--4253",
  month    =  may,
  year     =  1996,
}

@ARTICLE{Kirilyuk2010-ue,
  title     = "Ultrafast optical manipulation of magnetic order",
  author    = "Kirilyuk, Andrei and Kimel, Alexey V and Rasing, Theo",
  journal   = "Rev. Mod. Phys.",
  publisher = "American Physical Society",
  volume    =  82,
  number    =  3,
  pages     = "2731--2784",
  month     =  sep,
  year      =  2010
}

@ARTICLE{El_Hadri2017-zy,
  title     = "Materials and devices for all-optical helicity-dependent
               switching",
  author    = "El Hadri, Mohammed Salah and Hehn, Michel and Malinowski,
               Gr{\'e}gory and Mangin, St{\'e}phane",
  journal   = "J. Phys. D Appl. Phys.",
  publisher = "IOP Publishing",
  volume    =  50,
  number    =  13,
  pages     = "133002",
  month     =  mar,
  year      =  2017,
}

@ARTICLE{Shen2020-ml,
  title    = "Perpendicular magnetic tunnel junctions based on half-metallic
              {NiCo$_2$O$_4$}",
  author   = "Shen, Yufan and Kan, D and Lin, I and Chu, M and Suzuki, Ikumi
              and Shimakawa, Y",
  journal  = "Appl. Phys. Lett.",
  volume   =  117,
  pages    = "042408",
  month    =  jul,
  year     =  2020
}

@article{Xu2022-oi,
    author = {Xu, Xiaoshan and Mellinger, Corbyn and Cheng, Zhi Gang and Chen, Xuegang and Hong, Xia},
    title = "{Epitaxial {NiCo$_2$O$_4$} film as an emergent spintronic material: Magnetism and transport properties}",
    journal = {J. Appl. Phys.},
    volume = {132},
    number = {2},
    pages = {020901},
    year = {2022},
    month = {07},
    issn = {0021-8979},
    doi = {10.1063/5.0095326},
    url = {https://doi.org/10.1063/5.0095326},
}

@ARTICLE{Dho2022-bd,
  title     = "Magnetic domain structure of the ferrimagnetic (001) {NiCo$_2$O$_4$}
               film with perpendicular magnetic anisotropy",
  author    = "Dho, Joonghoe and Kim, Jungbae",
  journal   = "Thin Solid Films",
  publisher = "Elsevier",
  volume    =  756,
  pages     = "139361",
  month     =  aug,
  year      =  2022
}

@ARTICLE{Kan2020prb,
  title    = "Spin and Orbital Magnetic Moments in Perpendicularly Magnetized {Ni$_{1-x}$Co$_{2+y}$O$_{4-z}$} Epitaxial Thin Films: Effects of Site-Dependent Cation Valence States",
  author   = "Kan, Daisuke and Mizumaki, Masaichiro and Kitamura, Miho and Kotani, Yoshinori and Shen, Yufan and Suzuki, Ikumi and Horiba, Koji and Shimakawa, Yuichi",
  journal  = "Phys. Rev. B",
  volume   = 101,
  pages   = 224434,
  year     = 2020
}

@ARTICLE{Bitla2015,
  title    = "Origin of Metallic Behavior in {NiCo$_2$O$_4$} Ferrimagnet",
  author   = "Bitla, Yugandhar and Chin, Yi-Ying and Lin, Jheng-Cyuan and Van, Chien Nguyen and Liu, Ruirui and Zhu, Yuanmin and Liu, Heng-Jui and Zhan, Qian and Lin, Hong-Ji and Chen, Chien-Te and Chu, Ying-Hao and He, Qing",
  journal  = "Sci. Rep.",
  volume   = 5,
  pages    = "15201",
  year     = 2015
}

@ARTICLE{Kan2020-qo,
  title     = "Influence of deposition rate on magnetic properties of
               inverse-spinel {NiCo$_2$O$_4$} epitaxial thin films grown by pulsed
               laser deposition",
  author    = "Kan, Daisuke and Suzuki, Ikumi and Shimakawa, Yuichi",
  journal   = "Jpn. J. Appl. Phys.",
  publisher = "IOP Publishing",
  volume    =  59,
  number    =  11,
  pages     = "110905",
  month     =  oct,
  year      =  2020,
}

@article{Takahashi2021,
  title    = "Ultrafast Demagnetization in {NiCo$_2$O$_4$} Thin Films Probed by Time-Resolved Microscopy",
  author   = "Takahashi, Ryunosuke and Tani, Yoshiki and Abe, Hirotaka and Yamasaki, Minato and Suzuki, Ikumi and Kan, Daisuke and Shimakawa, Yuichi and Wadati, Hiroki",
  doi      = "10.1063/5.0058740",
  journal  = "Appl. Phys. Lett.",
  volume   = 119,
  number   = 10,
  pages    = "102404",
  year     = 2021
}

@ARTICLE{Shen2020,
  title    = "Tuning of Ferrimagnetism and Perpendicular Magnetic Anisotropy in {NiCo$_2$O$_4$} Epitaxial Films by the Cation Distribution",
  author   = "Shen, Yufan and Kan, Daisuke and Tan, Zhenhong and Wakabayashi, Yusuke and Shimakawa, Yuichi",
  journal  = "Phys. Rev. B",
  volume   = 101,
  number   = 9,
  pages    = "094412",
  year     = 2020
}

@article{Takahashi2023,
  title    = "Optically Induced Magnetization Switching in {NiCo$_2$O$_4$} Thin Films Using Ultrafast Lasers",
  author   = "Takahashi, Ryunosuke and Ohkochi, Takuo and Kan, Daisuke and Shimakawa, Yuichi and Wadati, Hiroki",
  journal  = "ACS Appl. Electron. Mater.",
  volume   = 5,
  number   = 2,
  pages    = "748--753",
  year     = 2023
}

\end{document}